# Quantitation of Cellular Dynamics in Growing *Arabidopsis* Roots with Light Sheet Microscopy

Giovanni Sena[1]*, Zak Frentz[2]*, Kenneth D. Birnbaum[1] & Stanislas Leibler[2,3]

[1] Department of Biology, Center for Genomics and Systems Biology, New York University, New York, NY, USA
[2] Laboratory of Living Matter, The Rockefeller University, New York, NY, USA
[3] The Simons Center for Systems Biology and School of Natural Sciences, Institute for Advanced Study, Princeton, NJ, USA

* The first two authors contributed equally to this work

## Abstract

To understand dynamic developmental processes, living tissues have to be imaged frequently and for extended periods of time. Root development is extensively studied at cellular resolution to understand basic mechanisms underlying pattern formation and maintenance in plants. Unfortunately, ensuring continuous specimen access, while preserving physiological conditions and preventing photo-damage, poses major barriers to measurements of cellular dynamics in growing organs such as plant roots.

We present a system that integrates optical sectioning through light sheet fluorescence microscopy with hydroponic culture that enables us to image, at cellular resolution, a vertically growing *Arabidopsis* root every few minutes and for several consecutive days. We describe novel automated routines to track the root tip as it grows, to track cellular nuclei and to identify cell divisions. We demonstrate the system's capabilities by collecting data on divisions and nuclear dynamics.

## Introduction

A broad understanding of development in both animals and plants requires a quantitative analysis of the morphological dynamics underlying tissue organization. The *Arabidopsis* root is a well established model system for studying pattern formation and organogenesis in plants [1]. It offers numerous advantages for live imaging, such as a high degree of transparency, radial symmetry, manageable size (typically 100-150 μm in diameter) and slow growth rate (few millimeters/day).

Live imaging techniques based on fluorescence confocal microscopy are used increasingly to study morphodynamics during post-embryonic development in plants [2,3]. Among these studies, high temporal resolution observations have been limited to short time-span events, as photo-induced cellular toxicity and fluorophore bleaching impose severe limitations on performing frequent time-lapse fluorescence microscopy over long periods [4].

However, relevant examples of plant organogenesis such as lateral organ formation [5] or organ regeneration [6] occur over periods of days. Many fundamental questions require long term analysis of numerous instances of individual developmental paths. For example, developmental robustness in pattern formation and maintenance may entail compensatory cell divisions or other morphological adjustments that span long periods. To capture long temporal correlations in cellular dynamics and to preserve the continuity of imaging data during the entire developmental process, extended observations of individual plants at high temporal resolution become necessary. To enable such an imaging approach, the sample must be continuously accessible for optical probing, while at the same time appropriate physiological conditions must be provided for the plant. While morphological analyses of *Arabidopsis* root growth have been carried out [7,8], these combined technical obstacles prevented a quantitative characterization at high spatiotemporal resolution and long temporal scale of cellular dynamics during root growth.

Light sheet microscopy (LSM) [9] offers numerous advantages over confocal microscopy for live imaging of thick tissues, such as reduced photo-bleaching and toxicity, and high speed acquisition [10]. At any given time during optical sectioning with LSM, the sample is illuminated by only a thin (<10 µm) laser "sheet", significantly reducing the corresponding incident photoenergy compared to the broad illumination of confocal microscopy. The need for lateral scanning is also eliminated, thereby reducing the duration of imaging. Recently, new techniques based on LSM have produced spectacular images of cellular dynamics at high temporal resolution during animal embryogenesis (in medaka, zebrafish and fruit fly) [11-13]. While still images of autofluorescence of isolated mulberry pollen grains have recently been recorded with LSM [14], this technique has not been implemented yet to achieve live imaging of growing multicellular plant structures.

We developed a new method specifically conceived to perform automated 3D fluorescence sectioning through LSM of a growing *Arabidopsis* root at high spatial (microns) and temporal (minutes) resolution, for many days. By taking advantage of the geometry and transparency of the root, the setup has been explicitly developed to acquire high quality data with minimal hardware complexity, resulting in a relatively inexpensive (~30,000 $) system composed of off-the-shelf optical and mechanical parts. Moreover, the computational analysis did not require special hardware and was performed on a single desktop computer.

Thus, our work comprises a simple and well-established imaging technique adapted to a novel growth and tracking system that can follow cellular dynamics in a continually growing organ.

## Results

### Frequent and long-term optical sectioning of the *Arabidopsis* root apex

In our system, frequent access to the sample is achieved by maintaining the plant in a micro-chamber mounted on the microscope's stage and designed to provide sterile conditions, continuous illumination, gas exchange and constant temperature. The root is confined to an optically accessible vertical channel in a glass fluorimetry cuvette and its continuous growth is sustained by a hydroponic perfusion system using commercial peristaltic pumps (Fig. 1A). In our implementation of LSM, an expanded laser beam is focused in one dimension by a cylindrical lens creating a thin (4.2 µm) excitation sheet used to optically section the root (Fig. 1B,C). The fluorescence signal is collected by a long working distance objective with its optical axis normal to the excitation sheet, as the whole root width is scanned by repeatedly moving the sample through the excitation sheet in steps of 1 to 2.5 µm in the axial ($z$) direction (Fig. 1B). An auto-focusing step precedes each acquisition, and scans are repeated every 10 minutes (details on the apparatus and the acquisition protocol can be found in the Methods). We detected no significant fluorophore photo-bleaching or photo-toxicity to the plant, as shown by the fluorescence intensity and rate of root growth over a typical experiment lasting more than 4 days (Fig. 1D).

Since the root typically grows by a few millimeters per day, the stage must be repositioned to keep the root apex in focus and within the field of view of the objective. Automatic sample tracking is achieved by optical feedback. The most recently acquired image is aligned to the previous one through global spatial translations and rotations and then the calculated translations are used to reposition the stage in three dimensions.

# Tracking cell nuclei

As a proof of principle, we present data obtained by imaging *Arabidopsis* transgenic plants expressing the protein fusion of the histone 2B (H2B) protein and the yellow fluorescent protein (YFP) under the control of the 35S promoter of the cauliflower mosaic virus [15] (Fig. 2A). The spatial expression pattern appears uniform with the exception of the cortex tissue, where a significant YFP depletion was observed and also confirmed by confocal microscopy.

Deconvolution based on a measured point spread function was applied to the stack of raw images acquired at each time point, achieving a final axial resolution < 4 μm across the imaged region. This resulted in the identification of most nuclei in the root, with an average inter-nuclear separation of 10 μm (Fig. 2A, S1). The temporal resolution of 10 minutes was sufficient to track nuclear separation during cytokinesis, which typically lasted 30 minutes, allowing the detection of cell divisions (Fig. 2B).

Fluorescence intensity collected from regions deep within the root along optical axes was attenuated, up to 60% along the collection axis and up to 35% along the light sheet axis. The pixel intensities were normalized along the two optical axes, and the images were segmented with a custom routine to extract the positions of nuclei at each time point. The segmentation routine performed globally with a 6.9% chance of missing true nuclei (false negative rate) and a 7.3% chance of erroneously labeling a nucleus (false positive rate), as calculated using a set of 1303 nuclei manually identified from a single time point. A more refined analysis of the error rates revealed variations in the segmentation performance in different regions of the root, but this did not prevent successful nuclear tracking throughout the root, as discussed below (Fig. 2C, Table S1). At each time point, the segmentation routine typically identified about 1500 nuclei.

We mapped the positions of all the nuclei in a moving frame of reference defined by the root tip and cylindrical coordinates were used to exploit the symmetry of the root (Fig. 3A). Local nuclear dynamics remained present in such a moving frame of reference, potentially reflecting a combination of sub-cellular nuclear motion, non-uniform stretching or twisting of the tissues and cell divisions.

As a first step to analyze the nature of the observed nuclear dynamics, we developed a method for automatically generating nuclear trajectories from the positions and morphological features of segmented nuclei. Tracking is a computational task commonly required in biology, for which several *ad hoc* methods have been developed [12,16-18]. Model based methods require estimation of parameters such as diffusion constants and probability densities for various motions, which are not trivial to measure independently [19]. The tracking problem can be generally formulated as a multidimensional assignment problem (MAP), facilitating the comparison of tracking data collected in different systems. This formulation is important for subsequent data analysis, especially for tracking populations of cells which may divide or are closely spaced relative to the typical displacement between time points, and for dealing with temporal gaps in the segmentation data for a given object due to false negatives. Given the computational complexity of MAP (NP complete [20]) and the large size of typical problems (more than 1000 objects for more than 100 time points in our case), exact solutions are not feasible, and rather than employ deterministic heuristics we used simulated annealing to efficiently search for approximate solutions [21]. A similar approach has been proposed and demonstrated by Gor *et al*. [22].

In our approach, a score or "energy" is defined for all possible assignments of nuclear identities, which we call configurations, so that correct tracking corresponds to low energy. The energy includes terms for the magnitude of nucleus displacement across neighboring time points, penalties for trajectories with temporal gaps and penalties for trajectories which fail to continue backwards or forwards in time or which branch. All terms were estimated empirically from acquired data; in particular, a library of manually identified cell divisions was used to define an energy term specific for cell division events. An initial configuration is generated by greedy local energy minimization, then the configuration is adjusted randomly and iteratively to reduce the energy until no further improvements are found.

We tested the performance of the tracking routine by manually verifying the consistency of nuclear assignments on a data set of 100 randomly picked trajectories, and observed a spatially uniform error rate of about 3% (Table S1).

# Nuclear collective motions: longitudinal stretching and rigid rotation of the root

We analyzed in detail a 29 hour interval from one of the acquisition sets. Visual inspection of the time-lapse movie of a deconvolved maximum intensity projection of the root (Video S1) and of single reconstructed trajectories (Fig. 3B) both suggested rich nuclear dynamics made up of collective motions and spatial fluctuations of individual nuclei. Fourier analysis of the displacements in several trajectories suggested a separation between low frequency directed motions and high frequency fluctuations (Fig. 3C). The predominant collective motion was clearly due to stretching along the root axis (Video S1), and was observed to vary over time.

To analyze collective motions, we first estimated the vector field of the nuclear velocities by binning spatial data on nuclear position and averaging over temporal windows (Fig. 4). The root's diameter at the most proximal observed region was found to decrease by about 10 μm during the 29 hour interval, with most of the change occurring between 8 and 21 hours (Fig. S2). We reasoned that such apparent contraction of the proximal region was simply due to cellular elongation progressively pushing the nuclei out of the field of view in the thickest, proximal, region. To separate temporal variations of velocities from this morphological effect, we divided the data set in two temporal windows: one including the first 21 hours and another including the remaining 8 hours.

Of the three velocity components, only that along the root axis was found to vary appreciably with time (Fig. 4A). Consistent with the radial symmetry of the root, no component of the velocity was found to have significant dependence on the cylindrical angle $\theta$ (not shown). Interestingly, we were able to detect a small, non-zero, average nuclear velocity along the cylindrical axis $m$ of the root starting around 90 μm from the root tip (Fig. 4A). Nuclei proximal to this point moved progressively away from the root tip with a constant average linear expansion rate in $m$ of $5.0 \times 10^{-4}$ (μm/min)/μm for the first 21 hours (Fig. 4A, early), declining to $2.3 \times 10^{-5}$ (μm/min)/μm for the last 8 hours (Fig. 4A, late), as shown by the linear fits of $v_m$ to $m$ in both cases. This reflected early longitudinal cell expansion.

The spatial average of the component of the velocity along the radial axis $\rho$ was indistinguishable from zero, and no spatial dependence was detected (Fig. 4B), ruling out both global and local cellular migration in the radial dimension. The angular component of the velocity showed a small but significant spatial average (0.017 μm /min) but no significant spatial dependence (Fig. 4C), indicating a rigid rotation but no detectable local twisting. Finally, we did not detect any statistically significant variation of either $v_m$ or $v_\theta$ with $\rho$, indicating no shear longitudinal or circumferential movements among tissues (Fig. 4).

## Identifying cell divisions

In the absence of cell migration, cell divisions play a central role in maintaining pattern organization in a growing root. We developed an automated routine to identify cell division events in the dataset of reconstructed nuclear trajectories (Fig. 5A). Manually identified cell divisions were found to have stereotypical dynamics: on average, the lateral size of the measured nuclear marker decreased by 20% following division, while the intensity increased as much as 5 fold immediately prior to division, followed by a rapid decrease to below the temporal mean intensity and a reversion to the mean intensity over the 30 minute period following division. These characteristic dynamical features were used, after the nucleus tracking, to automatically separate true divisions from spurious events due to optical artifacts or segmentation errors. Overall, our method performed with a 26% chance of missing true divisions (false negative rate) and 13% chance of erroneously labeling an event as division (false positive rate; Fig. S3).

## Distribution of cell divisions and their orientations

The temporal distribution of all detected cell divisions was inconsistent with a Poisson process with constant rate, indicating non-trivial temporal correlations (Fig. 5B). The cell divisions appeared mainly clustered in a region between 100 to 200 µm from the root tip (Fig. 5C), making our large dataset a unique tool for the future characterization of this region of proliferation in the root meristem.

The orientation of a cell division determines its effect on the topology of tissue organization. The spatial and temporal resolutions of our observations allowed the analysis of cytokinesis: the displacement of daughter nuclei from the parent position over one time interval defines the orientation of the cell division. For example, a longitudinal cell division is defined as having its orientation along the main axis $m$ of the root. In our data daughter nucleus separation occurred symmetrically (Fig. 2B and 5A) and tended to be larger by a factor 2.4 than the typical displacement over the same interval for non-dividing nuclei. In the temporal interval we examined, most of the detected divisions occurred approximately parallel to the main axis $m$ (Fig. 5C,D, red crosses), with only a few occurring along the circumferential and radial dimensions (Fig. 5C,D, blue and green crosses, respectively). This is consistent with a scenario where most of the cellular proliferation contributes to root longitudinal growth, with only few events resulting in radial tissue "invasions" or perturbing the circumferential organization of single tissues. Due to its resolution and the long duration of the observations, our system is thus ideal for a systematic measurement of the orientations in 3D of cell division events in a growing root.

# Discussion

The presented apparatus represents a unique tool for supporting a quantitative approach to plant morphodynamics.

As a proof of principle, we present data collected from *Arabidopsis* roots expressing a nuclear localized fluorescent marker, producing a detailed analysis of the nuclear dynamics during post-embryonic root growth. The high temporal resolution of the imaging was sufficient to track densely packed nuclei, to detect short scale fluctuations in their position (interpreted as sub-cellular erratic movements of the nuclei) as well as large scale directed motions and to identify division events. The sensitivity of our method is revealed by the small average (collective) positive nuclear velocity along the longitudinal dimension *m* that we were able to detect just 90 µm from the root tip. We interpret the constant increase of this quantity as the well known longitudinal stretch of the tissues, caused by cellular elongation along the main axis of the root. Besides an expected rigid rotation of the root around its main axis *m*, we did not detect any significant changes of the nuclear average angular velocity along *m*, indicating that no or very little twisting of the tissue occurred in the region analyzed. Moreover, as largely expected in a structure where cell walls produce a strong anchoring effect, we did not find evidence for cell migrations. Finally, the non-uniform temporal distribution observed for the cell division events, together with the decrease observed in the longitudinal expansion rate after 21 hours in the data set analyzed, suggests physiological transitions during root growth in the experiment's conditions. These observed changes could simply reflect a combination of aging and metabolic adaptation to the hydroponic conditions as the root is transferred from a solid growth medium (standard MS agar plate) to the continuous perfusion of liquid medium (new imaging and growth micro-chamber). Even though we cannot exclude that the imaging procedure itself contributes to physiological changes in the root, the typical growth rate indicates no major photo-toxic effects.

The problem of tracking populations of cells has arisen in many fields and for several methods of microscopy. The general formulation of the tracking problem as a multiple assignment and the use of stochastic optimization to find approximate solutions may be useful in studies of organisms other than *Arabidopsis* and for types of microscopy besides light sheet based methods. Though the energy defined for our nucleus tracking algorithm was specific to the organism, microscope and fluorescent marker, all terms were estimated directly from observations, a process which should readily extend to other systems.

The versatility of the method allows not only for the visualization of the root 3D structure by identifying cell positions, as shown here, but also potentially for tracking dynamics of endogenous gene expression in different transgenic lines. Furthermore, to capture the kinetics of *in vivo* cellular trans-differentiation, additional fluorescence channels could be added to simultaneously image multiple cell type-specific reporters. This capability opens the potential to locate and quantify symmetric and asymmetric divisions, a critical issue in development.

The ability to produce frequent time-lapse imaging of the 3D structure of the *Arabidopsis* root at cellular resolution as it grows vertically for many days, without inducing significant photo-bleaching or photo-damage, offers a novel dynamic perspective on root development. The resulting quantitative phenomenology will contribute, for example by supporting mathematical modeling of organogenesis, to our understanding of many fundamental developmental processes in plants. Among these are pattern maintenance in presence of continuous post-embryonic tissue growth and *de novo* pattern establishment during organ branching and tissue regeneration. From an even broader perspective, the relatively inexpensive setup presented here could be used in replicate to produce a rich quantitative characterization of the effects of pattern perturbations induced by genetic mutations, organic or inorganic molecules such as plant hormones or poisoning agents, and biotic or abiotic stresses in general.

# Methods

## Plant imaging and growing chamber

For each experiment a plantlet was grown in a closed and sterile micro-chamber (Fig. 1A) designed to constrain the root growth to an optically transparent vertical narrow path, while at the same time providing illumination and gas exchange for the leaves. The main body of the chamber is made up of a glass fluorimetry cuvette (inner size 10 x 10 x 40 mm, Starna) with a glass capillary tube (o.d. 1.5 mm, Fisher) placed vertically against one corner of the cuvette to create a channel (diameter approximately 300 μm) in which the root grows. Two teflon collars with a thickness of 80 μm are added at the ends of the tube to distance it from the walls of the cuvette and create enough space for medium flow in the channel, while still preventing root escape. The tube is held in position by filling the cuvette outside the channel with 3 mm glass beads (Fisher), and the whole space is filled with liquid medium (see Perfusion, below). The cuvette is closed with a custom-made transparent top cap and the junction sealed with several rounds of Parafilm. The cap is obtained from a disposable plastic spectrophotometer cuvette (inner size 10 x 10 x 40 mm, Fisher) whose square base has been removed and whose three long sides have been cut to create windows. The base is then closed with a glass square coverslip (for illumination) glued with silicone and the three side windows with adhesive breathable sealing tape (for gas exchange, Nunc).

The sample is illuminated from above by an array of 12 bright white LEDs (Luxeon Star/O) powered by a current source controlled by the computer. The LED current is gated by a signal indicating camera exposure so that the LEDs are shut off during exposure. The optics and micro-chamber are housed in a thermally insulated light-tight box with a hinged door allowing sample loading and a thermoelectric assembly (AC-046, TE Technologies, Inc.) maintains constant temperature within the box.

## Perfusion

To sustain root growth for days, continuous perfusion of liquid medium is maintained through the micro-chamber. One peristaltic pump (Dynamax RP-1, Rainin, run at 30 rpm with Pharmed BPT tubing 0.51mm i.d., Cole Parmer) flows liquid medium from a reservoir to the bottom of the micro-chamber, and a second pump (Dynamax RP-1, Rainin, run at 30 rpm with Pharmed BPT tubing 0.89 mm i.d., Cole Parmer) flows medium out of the top of the glass cuvette (right below plantlet's hypocotyl) to a waste bottle. The outflow rate is kept higher than the inflow rate (different i.d.), to prevent flooding. The liquid medium is composed by 1/4X Murashige and Skoog (MS) basal medium enriched with Gamborg's vitamins (Sigma), 3.0% sucrose, 0.05% MES buffer (Sigma) and adjusted to pH 5.7 (KOH). After autoclave sterilization, antibiotic antimycotic solution (A5955, 1/200X, Sigma) and Amphotericin B (A9528, 1.0 mg/l, Sigma) are added the medium. The medium reservoir is maintained saturated with oxygen with filtered air from a standard aquarium air pump.

## Plant material and growth conditions

*Arabidopsis* seeds are sterilized as describe previously [6] and germinated on 0.8% agar plates containing 1X MS, 0.5% sucrose, 0.05% MES buffer, pH 5.7, under constant illumination and 23°C. Between 3 and 5 days post-germination a single plantlet is transferred to the micro-chamber, first making sure that the root is completely aligned along the corner of the cuvette, and then reconstructing the micro-chamber by adding the capillary tube, filling the cuvette with the glass beads, adding some liquid medium and sealing the top cap with Parafilm. The perfusion is started, and the automated tracking and scanning sequence initiated. Throughout the experiment, the micro-chamber is illuminated with constant light at 20,000 lux and the temperature maintained constant at 23°C.

## Optical apparatus

Our imaging system is based on the light sheet microscopy (LSM) concept [9]. Fluorescence excitation is provided by either a 473 nm, 80 mW diode pumped solid state laser (Lasever) or a multi-line 60 mW Argon ion laser (Spectra Physics). The laser beam passes an excitation filter (Chroma) and optional neutral density filter (Thorlabs) and is modulated by a Pockels cell (350-50, Conoptics). Two flat mirrors direct the beam into a 10X Galilean beam expander (Thorlabs) illuminating a circular 9 mm iris (Thorlabs). The circular beam is focused in one dimension by a plano-convex cylindrical lens (f=30 mm, Thorlabs), creating a "sheet" of 4.2±0.2 µm across 180 µm along the excitation axis. Emission is imaged with a 20X super long working distance objective (NA=0.35, f=19.9 mm, Nikon) coupled by a lens tube to an emission filter (Chroma) and cooled CCD camera (1392 x 1024, 4.65 µm square pixels, QICam, QImaging). Acquisitions are typically performed with 2 x 2 pixel binning.

The $x$ axis is defined as coincident with the optical axis of the cylindrical lens, the $y$ axis is parallel to gravity and $z$ is the optical axis of the objective (Fig. 1B).

The sample is maintained in the coincident foci of the cylindrical lens and the objective by 3 translation stages coupled to the sample holder ($x,y$ and $z$ axes, Fig. 1B) and 2 translation stages coupled to the CCD camera and objective ($x$ and $z$ axes). All translation stages are positioned by micrometers driven by servo motors (25 mm travel, 40 nm resolution).

## Resolution

The axial resolution of the optical system is determined by the sheet thickness near its focus and the depth of field of the microscope objective. Since the former is inversely proportional to the numerical aperture of the cylindrical lens (0.15) and the latter is inversely proportional to the square of the objective's NA ($0.35^2 = 0.1225$), the sheet thickness dominates in determining the overall axial resolution. The mean of the sheet full width at half maximum (FWHM) over the region of interest is 4.2±0.2 µm, as measured by scanning fluorescent beads (0.3 µm diameter, Duke Sci.) embedded in Phytagel (0.5%). This sheet thickness is maintained within 5% over 180 µm along the excitation beam axis, which is sufficient to section the entire diameter of the root. The subsequent deconvolution step (see below) improves the axial resolution, as shown by the intensity profiles of neighbor nuclei (Fig. S1).

## Image acquisition

At the beginning of each experiment, the root tip is positioned in the focus of the excitation beam, whose profile was previously measured by scanning fluorescent beads as described above. Imaging typically occurred every 10 minutes. At the beginning of each scan, the focal plane of the objective is determined by exciting fluorescence from a central region of the root while imaging at several positions along the optical axis of the objective. At each position, the spatial gradient is estimated at each pixel and summed over the image, and the position corresponding to the maximum sum is chosen as the best focus for that excitation plane. During the scan, the sample is repeatedly moved through the laser sheet by stepping its stage in the $z$ direction and acquiring the emission signal from each optical section (typical exposure time was 25-50 ms). At each step of the scan, the stage holding the camera and objective is repositioned in $z$ to maintain the excitation plane in the objective focus, accounting for the defocus caused by variable optical path length through air and medium.

The sectioning interval is chosen to satisfy the Nyquist criterion for the measured point spread function (typically 1.25 - 2.5 µm). The total scan depth (typically 200 µm) is larger than the root thickness (typically 100 µm) to ensure that the entire tip is sectioned. The camera acquisition triggers via TTL a pulse from a custom circuit to modulate the Pockels cell. Typically multiple acquisitions at each $z$ step are averaged to reduce amplifier noise from the CCD.

The duration of the scan (typically 90 s) is primarily limited by the repositioning of the motorized stages.

# Photoenergy

Due to the fundamental difference between the way the optical sectioning is performed in LSM and in confocal laser scanning microscopy (CLSM), a scan consisting of $n$ slices (for typical *Arabidopsis* root, n ≈ 60 with 2.5 µm steps) will produce a total incident photoenergy at least $n$ times reduced in LSM compared to CLSM. Moreover, the calculated irradiance in the focal plane is 57 W/cm$^2$ with LSM compared to 950 kW/cm$^2$ calculated for typical confocal imaging of a root expressing the same marker. Such 16,000-fold reduction in irradiance and 60-fold reduction of incident photoenergy for the LSM compared to CLSM allow acquisition of live plant roots at high temporal resolution without significant photo-bleaching or photo-toxicity, as seen in Figure 1D of the main text.

# Root tracking

Because the root grows continuously throughout the experiment (approximately 1-3 µm/min) the sample must be repositioned between scans to maintain the root tip in the field of view. This is accomplished via optical registration of the most recent scan's images to those from the previously optically registered scan. The algorithm used for registration is a version of the Lucas-Kanade algorithm [23]. The algorithm calculates by iterated gradient descent the 3D rigid motion which minimizes the squared difference between two images, assuming that the motion is small compared to the length scale over which the image intensity varies. After the optimal rigid motion has been calculated, the stages are translated to reposition a point of interest in the most recently scanned image to a fixed location in image space, chosen to maintain the entire root tip within the field of view. The stage position is recorded at each scan.

The root typically rotates about all three axes as it grows. The calculated rigid motion includes rotations, while the repositioning of the sample stage only involves translations - the former are necessary to maintain registration of the root tip. Tracking was accurate for the largest observed translations, which were 15% in the linear dimension of the field of view.

Defocus along the excitation beam caused by movement along the $x$ axis of the sample was compensated taking into account the refractive indices of air and medium (water).

# Acquisition software

The acquisition and tracking is facilitated by custom C++ software with a graphical interface controlling the stages, camera, LEDs, perfusion and all scan parameters, and displaying relevant images and information from the acquisition as it proceeds. The most computationally demanding task during an acquisition is the sample tracking, which takes about one minute on a desktop computer with a dual core 3.0 GHz processor (Core 2 Duo E8400, Intel) and 2GB RAM.

# Deconvolution

The stack at each time point is deconvolved off-line via expectation maximization (EM) assuming Poisson noise and using the point spread function measured by imaging 0.3 μm diameter fluorescent beads (Duke Scientific Corp.) [24]. Typically 30 iterations of the EM algorithm are used. 3D fast Fourier transforms were performed using FFTW [25]. The deconvolution significantly improves the separability of nuclei along the optical axis (Fig. S1), without introducing artifacts affecting the analysis.

# Segmentation

The stack of deconvolved images at each time point is automatically segmented by a routine adapted from a previously described method [12]. The main difficulties for image segmentation in our data are heterogeneity of nuclear size and intensity, and variation of optical quality with position within the root. In addition, the spatial extent of each nucleus appears comparable to the internuclear distance, occasionally making the separation of neighboring nuclei even more challenging.

Most of the spatial variation can be ascribed to the depth of tissue through which the light passes along the two optical axes. These depths are estimated from images by constructing the three dimensional convex hull of all voxels above an intensity threshold chosen to separate nuclear signal from background. This convex hull approximates the boundary of root tissue and for each imaged voxel the distances along the optical axes to the hull can be calculated. The logarithm of maximum pixel intensity was fit to a second order polynomial in the two optical depths by least squares and the fit was subsequently used to normalize intensities within the root.

A segmentation routine was developed consisting of an increasing series of global segmentation thresholds and a set of criteria for classifying a contiguous blob as a nucleus. Once a blob is classified as a nucleus, its voxels are no longer considered as the segmentation threshold increases. The criteria for classification are ranges for number of voxels, mean intensity, lateral size, lateral eccentricity and axial size. To ensure that nuclei are separated along the optical axis, the intensity profile along the optical axis is computed for each blob and tested for the absence of multiple maxima as a further criterion. A blob must fall within the specified range for each parameter to be classified as a nucleus. The ranges for classification resulted from training with 850 manually identified nuclei, located throughout the root tip

The output of the segmentation routine is a time series of centroid location, maximum and mean intensities, voxel count, lateral size, lateral eccentricity and axial size for each detected nucleus.

The segmentation validation was performed on a set of 1303 manually identified nuclei. The routine failed to identify 90 of the nuclei in the test set (false negative rate, 6.9%), while labeling as nuclei 95 blobs not present in the test set (false positive rate, 7.3%) (Table S1).

# Registration of segmented images and frame of reference

The global alignment of nuclear positions over time is important, as the nuclear displacement across time points is the essential factor used in tracking nuclei. The sample tracking is predictive, in that it compares the previous two stacks in order to calculate a translation for the next scan. An improved alignment with finer translations and rotations can be calculated after the image has been acquired. For simplicity, we calculate this alignment based on nuclear positions extracted by segmentation, minimizing by iterated gradient descent the mean squared nuclear displacements over adjacent time points with respect to rotations and translations. This frame of reference is used for nuclear tracking.

Finally, a new frame of reference is introduced in which the origin of the primary root axis $m$ effectively coincides with the root tip, defined at a constant distance from a single nucleus identifiable in all time points, and the cylindrical angle $\theta$ is measured from the objective axis (Fig. 3A). All the data presented are in such reference frame, which facilitates the analysis of performance as a function of penetration depth.

# Nuclear tracking

To follow nuclei through time and identify cell divisions, the output of the segmentation routine is analyzed by a custom algorithm for nuclear tracking. Our routine exploits similarity in nuclear position within the root from one time point to the next, since other features such as intensity and nuclear size fluctuate more than position due to biological and optical effects.

Nuclear identity assignments are represented as an array of trees, with each tree corresponding to a unique identity, each node of a tree representing a segmented blob associated with that tree's nuclear identity, and an edge between two nodes indicating that the corresponding blobs represent the same nucleus at adjacent times. In the ideal case, each level of the tree corresponds to one scan of the experiment, but in practice false negatives in the image analysis necessitate skipping isolated time points for some nuclei, so that a nucleus at time $t$ can be directly associated with a nearby nucleus not just at time $t+1$ but also $t+2$, etc.

The energy to be minimized consists of a sum over all edges of all trees of squared spatial displacement, a penalty for temporal gaps, a penalty for terminal nodes and a penalty for branching. Each identified nucleus is numbered and for nucleus $i$ the position $r_i$, time $t_i$ and maximum intensity $I_i$ are considered. An indicator function $A_{ij}$ is defined as equal to 1 when nucleus $i$ is directly equated with nucleus $j$ and 0 otherwise. The energy is expressed in $\mu m^2$, since the squared displacement is the most important term and is the only term with physical meaning, so this scale eases interpretation of the various terms.

The energy can be written as

$$\sum_i f(m_i) + g(n_i) + \sum_{j:t_j<t_i} A_{ij}\left(\left|r_i - r_j\right|^2 + \alpha(t_i - t_j - 1)^2\right) \quad (1)$$

where $\alpha = 20\,\mu m^2$ is a coefficient controlling the contributions from non-physical quantities, $m_i$ and $n_i$ are the number of nuclei identified with nucleus $i$ at immediately previous and later times along its trajectory, respectively:

$$m_i = \sum_{j:t_j<t_i} A_{ij}$$

$$n_i = \sum_{j:t_j>t_i} A_{ij}$$

and $f(m)$ and $g(n)$ are functions defined as follows for penalizing nuclei with numbers of parents and children not equal to one, respectively. Although nuclear fusions ($m_i = 2$) do not occur biologically, they are allowed in the analysis with a finite penalty to prevent false positives from disrupting the continuity of nearby trajectories so that these events can be considered manually:

$$f(m_i) = \begin{cases} 100\,\mu m^2 & if \quad m_i = 0 \\ 0 & if \quad m_i = 1 \\ 195\,\mu m^2 & if \quad m_i = 2 \\ \infty & if \quad m_i > 2 \end{cases}$$

The data contained many events in which two nearby nuclei were properly resolved by the segmentation routine for some time, followed by a period in which they were not resolved. When the nuclei are again resolved, the event resembles a cell division. In order to distinguish these events from true cell divisions, several properties unique to true divisions were incorporated in the branching penalty.

In a manually generated set of over 400 true divisions, the displacement of centroids from the parent cell to each daughter cell was on average 2.4-fold greater than the average non-dividing displacement (2.6 ± 0.6 µm for dividing cells compared to 1.1 ± 0.8 µm for non-dividing cells over 10 minute scan intervals). In addition the displacements from the parent were typically in opposite directions.

The branching term g(0) was designed to apply negligible penalty to true division events and significant penalty to most events which otherwise resembled divisions, based on these empirical observations. For a nucleus $i$ with two daughter nuclei $j$ and $k$, define the displacements $\Delta r_{ij} = r_i - r_j$ and $\Delta r_{ik} = r_i - r_k$ and the deviations of these displacements from the average measured displacement upon division $\delta_{ij} = |\Delta r_{ij}| - \langle |\Delta r| \rangle_{divisions}$, where $\langle |\Delta r| \rangle_{divisions} = 2.6\,\mu m$, along with a threshold deviation: $\delta_{th} = 1.25\,\mu m$.

The penalty for node $i$ having $n_i$ children is defined as

$$g(n_i) = \begin{cases} 100\,\mu m^2 & if \quad n_i = 0 \\ 0 & if \quad n_i = 1 \\ L_{ij} + L_{ik} + \omega_{ijk} & if \quad n_i = 2 \\ \infty & if \quad n_i > 2 \end{cases}$$

where $L_{ab}$ accounts for the displacement of the daughter nucleus from the parent:

$$L_{ab} = \left(1 - rect\left(\frac{\delta_{ab}}{2\delta_{th}}\right)\right)\left(1.5\,\mu m^2 + (\delta_{ab} - \delta_{th})^2\right)$$

with

$$rect(x) = \begin{cases} 1 & if \quad |x| < 1/2 \\ 1/2 & if \quad |x| = 1/2 \\ 0 & if \quad |x| > 1/2 \end{cases}$$

and $\omega_{abc}$ accounts for the angle between the two displacements:

$$\omega_{abc} = 75 \mu m^2 \left(1 + \frac{\Delta r_{ab} \cdot \Delta r_{ac}}{|\Delta r_{ab}||\Delta r_{ac}|}\right)$$

The fundamental term in the energy is the squared displacement, since nuclear position is the strongest indicator of identity in our data. The penalties $f(0)$ and $g(0)$ for zero children and parents determine the length scales over which identifications can be made, as described below. The remaining terms are chosen to cope with specific events - specifically, segmentation errors and cell divisions.

The values for the penalties are empirically constrained. As an example, we depicted in Figure S4 a typical situation: two nuclei, one at time $t$ and one at time $t+1$, are separated by a distance $\Delta r$. Consider two configurations: (A), in which the nuclei are connected, i.e. assigned to the same trajectory, and (B), in which the nucleus at time $t$ has no child and the nucleus at time $t+1$ has no parent, i.e. not assigned to the same trajectory. From (1), the energy for (A) is $(\Delta r)^2$, while the energy for (B) is $f(0) + g(0)$. Configuration (A) should be chosen to describe a situation where the same nucleus is observed at both time points with a true displacement $\Delta r$, while configuration (B) describes events in which the first nucleus disappears after time $t$ (by leaving the field of view, for example) and the nucleus observed at time $t+1$ is its neighbor. Since the algorithm favors lower energy configurations, to achieve the correct assignment $f(0) + g(0)$ must be large compared to the square of the largest observed nuclear displacement over the time-step ($d_{max} = 5$ μm), so that the energy of (B) is always higher than the one of (A) in the case of true displacement of a single nucleus. At the same time, $f(0) + g(0)$ cannot be much larger than the square of the typically observed internuclear distance ($d_{nn} = 10$ μm), otherwise distinct neighbor nuclei will tend to be wrongly assigned to the same trajectory as in configuration (A).

Other parameters are determined by similar empirical considerations. The tracking parameters have not been systematically optimized; nevertheless their performance allows for useful analysis, as validated below.

The scheme for optimizing identity assignments was as follows for all experiments. The initial configuration is chosen by equating each nucleus at time $t$ with its nearest neighbor at time $t+1$. The move at each step is chosen randomly but with a bias for high energy nodes as targets, so that the probability of choosing a given nucleus for a move is proportional to the energy it contributes to the total. Once node $i$ has been chosen, the possible moves include cutting a connection by flipping $A_{ij}$ from 1 to 0 for some $j$, adding a connection by flipping $A_{ij}$ from 0 to 1, or changing a connection by flipping $A_{ij}$ to 0 and $A_{ik}$ to 1 for some $j$ and $k$. Moves are accepted according to the Metropolis criterion [26]. We found that allowing two serial moves, where one of the nodes involved in the first move is chosen as the target for the second move, improved the rate of convergence significantly, while allowing three or more such serial moves showed no improvement.

We found that the critical temperature for annealing the energy defined in equation (1) was approximately $50 \mu m^2$. This was typically chosen as the initial temperature, the system was equilibrated for several thousand moves, then exponentially cooled.

Nuclear tracking was manually validated by choosing 100 trajectories at random and identifying errors by considering high energy nodes and comparing to images when necessary. Since some ambiguity remains in the manual tracking, we calculated global lower and upper bounds on the error rate of 2.4% and 3.9%, respectively, over 32,067 total assignments (Table S1). Nuclear tracking was performed on a dual core 2.66 GHz processor (Core 2 Duo E7300, Intel).

## Identification of cell divisions (or cell-lineage tracking)

Each branching event in the set of identified nuclear trajectories was labeled as a potential cell division. The majority of these branching events arose from failure of the segmentation to properly separate two nearby nuclei at one time point followed by proper separation at subsequent time points. Additional dynamic features of true cell divisions were considered after the nucleus tracking was performed. These features did not enter the definition of the energy for tracking due to the expense of calculation arising from their temporally extended nature. This filtering step was implemented by constructing a support vector machine [27] (SVM) for classification based on nuclear size and intensity at time points within a temporal neighborhood of the identified division and on the durations of the trajectories for the parent nucleus and both daughter nuclei. A complete set of 461 true cell divisions was manually built, and the SVM was trained on a subset of 424 true cell divisions and 3805 spurious branchings, yielding an out of sample accuracy of 92.3% in five fold cross validation. The SVM was then used to classify all branching events and compared to the complete manual set, resulting in 340 true positives (73.8% of true divisions) and 58 false positives (12.6% of true divisions).

## Source code

Code for deconvolution, segmentation and nucleus tracking is available and maintained as a SourceForge project ([qanneal-celltrk.sourceforge.net](qanneal-celltrk.sourceforge.net)).

## Acknowledgements

We thank B. Scheres (Utrecht University) for providing seeds of *Arabidopsis* transgenic plants expressing *35S::H2B::YFP*. We thank D. Hekstra (The Rockefeller University) for advices in building the optical system, and J.Chuang, D. Hekstra, S. Kuehn (The Rockefeller University) and T. Nawy (New York University) for helpful comments on the manuscript.

# Figure Legends

**Figure 1 | Imaging system**. (**A**) Plant micro-chamber details. Left panel, view from the top; Right panel, view from the side showing inflow and outflow of perfusion medium. B, beads; CT, capillary tube; Ch, channel; Co, collar. (**B**) Optics. L, laser; F1, excitation filter; PC, Pockels cell; BE, beam expander; I, iris; CL, cylindrical lens; IGC, plant imaging and growing micro-chamber on a stage moving in *xyz*; O, objective; F2, emission filter; C, camera on a stage moving in *xz*. (**C**) 3D rendering of the laser sheet scanning the corner of the cuvette where the root is growing; P, plant (**D**) Long-term sustained root growth. Root tip speed (blue line) as measured during tracking in a typical experiment; average fluorescent intensity normalized to the beginning of experiment, detected from the same root expressing *35S::H2B::YFP* (green line).

**Figure 2 | Deconvolution and segmentation**. (**A**) *Arabidopsis* root expressing *35S::H2B::YFP*. Maximum intensity projections before (left panel) and after (right panel) deconvolution; the horizontal axis corresponds to the objective axial dimension; bar, 50 µm. (**B**) Cell divisions from same root in (A), time-lapse. Three cells divide giving rise to daughters labeled with "a" and "b"; bar, 5 µm. (**C**) Density plot of segmented nuclei. Projection along the longitudinal axis of all the nuclei recognized by the segmentation, accumulated in a 29 hour period. Five regions used in Table S1: I, towards objective; II, away from laser source; III, away from objective; IV towards laser source; V center of the root ($\rho < 30$µm).

**Figure 3 | Trajectories**. (**A**) Cylindrical coordinates. Schematic of the root tip and the cylindrical coordinate system used in text: *m*, distance from the tip on the main longitudinal axis of the root; $\rho$, distance from the center of the root; $\theta$, angle with respect to the axis normal to *m* and with maximal component *z* (see Fig. 1B), where $\theta = 0$ towards the objective and $\theta = \pi/2$ towards the laser source. (**B**) Nuclear displacements along a typical trajectory, displayed along the main axis of the root (*m*, top panel), the *m*-normal closest to *z* ($\rho cos\theta$, middle panel) and the *m*-normal closest to *x* ($\rho sin\theta$, bottom panel); bar, 10 µm. (**C**) High frequency (left panel, freq.>0.23 hr$^{-1}$; bar, 1 µm) and low frequency (right panel, freq.<0.23 hr$^{-1}$; bar, 20 µm) components of the trajectory shown in (B).

**Figure 4** | **Analysis of collective nuclear velocities**. Heatmaps and plots for each cylindrical component of the velocity, from data collected in a 29 hour temporal interval. In the heatmaps, only bins containing 36 or more nuclei are shown; in the plots, the solid black line is the median per 5 micron bin, while the dashed black lines are the 25% and 75% quartiles. (**A**) *m* component of the velocity; "early", temporal window 0-21 hours; "late", temporal window 21-29 hours; the red dotted line is the linear fit of $v_m$ for the region 80-300 µm. (**B**) $\rho$ component of the velocity. (**C**) $\theta$ component of the velocity multiplied by the position $\rho$ in the radial dimension.

**Figure 5 | Cell divisions**. (**A**) Nuclear displacements along a typical trajectory showing branching, displayed along the main axis of the root (*m*, top panel), m-normal closest to *z* ($\rho cos\theta$, middle panel) and the *m*-normal closest to *x* ($\rho sin\theta$, bottom panel); arrow, branching point indicating cell division; bars, 20 µm. (**B**) Temporal distribution of the cell divisions detected in the 29 hour interval analyzed (true positives, N=340). (**C**) Position and orientation of the same cell divisions, superimposed to a smoothed density estimate of all the nuclei segmented at all time points throughout the same 29 hour interval; main direction of divisions: red, $\Delta m$; blue, $\rho\Delta\theta$;

green, *Δρ* (**D**) Polar histograms for the angles of the cell division orientation to *Δm*, in the plane defined by *Δm* and *Δρ* (top panel) and by *Δm* and *ρΔθ* (bottom panel); color code as in (C).

## Supporting Information

**Figure S1 | Effects of deconvolution in the axial (*z*) dimension**. (**A**) Raw (left column) and deconvolved (right column) images of three close nuclei, at different *z*-positions; bar, 7.5 μm. (**B**) Maximum intensity collected in the region marked by the red ellipses in (A), both for the raw (black) and the deconvolved (blue) images. The increased resolution in *z* after deconvolution is shown by the improved separation between peaks.

**Figure S2 | Contraction of the root radius.** Number of nuclei with radial coordinate *ρ* greater than 55 μm, as a function of time. The red dotted line indicates the time point (21 hours) chosen to distinguish "early" and "late" temporal windows.

**Figure S3 | False positives**. Positions of the false positive cell divisions found in the same dataset containing the true positives shown in Fig. 5 of the main text, superimposed on the smoothed density estimate of all the nuclei segmented at all time points throughout the same 29 hour interval.

**Figure S4 | Tracking nuclei: example.** Diagram showing the correct assignments in two typical scenarios (A and B) for two consecutive time-points (*t* and *t*+1). (**A**) The same nucleus **a** is observed at both times *t* and *t*+1, with a spatial displacement (blue arrow): the two nuclei should be assigned to the same trajectory (link between them, red arrow). (**B**) Nucleus **a** is only observed at time *t* and nucleus **b** is only observed at time *t*+1: the two nuclei should not be assigned to the same trajectory (no link between them, red truncated links). *Δr*, distance between nuclei observed in *t* and *t*+1; $d_{max}$ = 5 μm, maximum nucleus displacement observed between t and t+1 (time-step = 10 min); $d_{nn}$ = 10 μm, average distance between nearest neighbor nuclei.

**Table S1 | Performance of segmentation and tracking routines.** Segmentation: error rates in nuclei recognition in the regions I-V defined in figure 2C; N, number of manually identified nuclei; FP, false positive rate; FN, false negative rate. Tracking: error rates in nucleus-trajectory assignment in the same regions I-V; N, number of manually identified trajectories starting in the given region; error range, lower and upper bounds for errors in assigning a nucleus to the correct trajectory.

**Video S1| Growing root**. Time-lapse images of the same 29 hour interval used to extract the presented data. Maximum intensity projection along the optical axis of the deconvolved images. The temporal step is 10 minutes. Longitudinal tissue stretching and several cell divisions are detectable by eye.

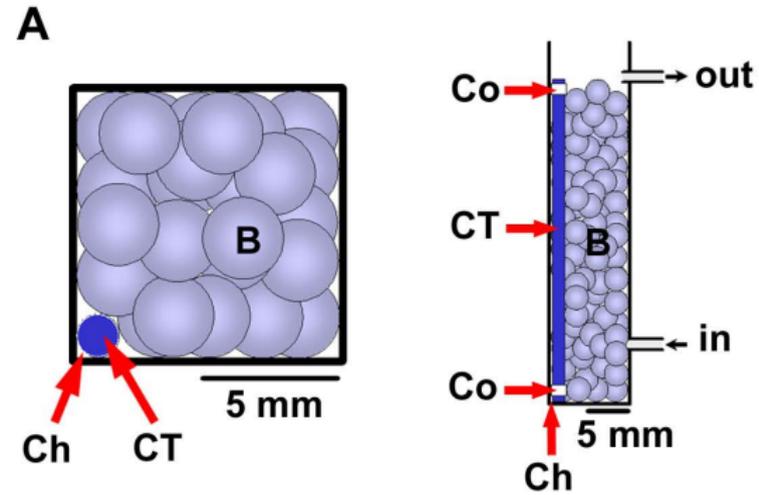
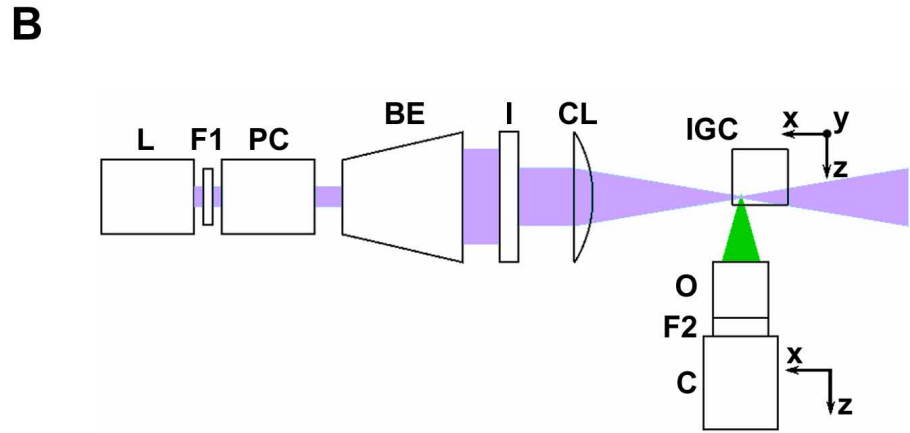
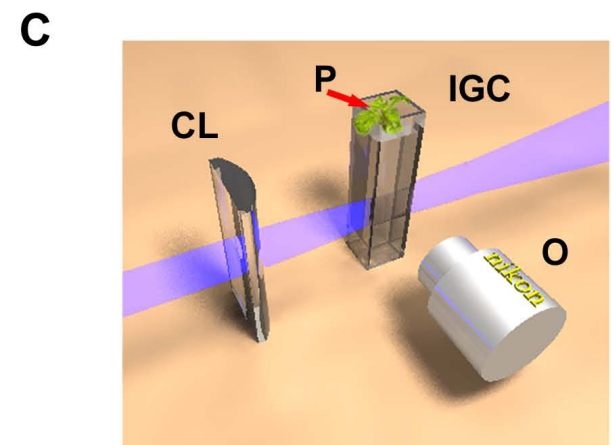
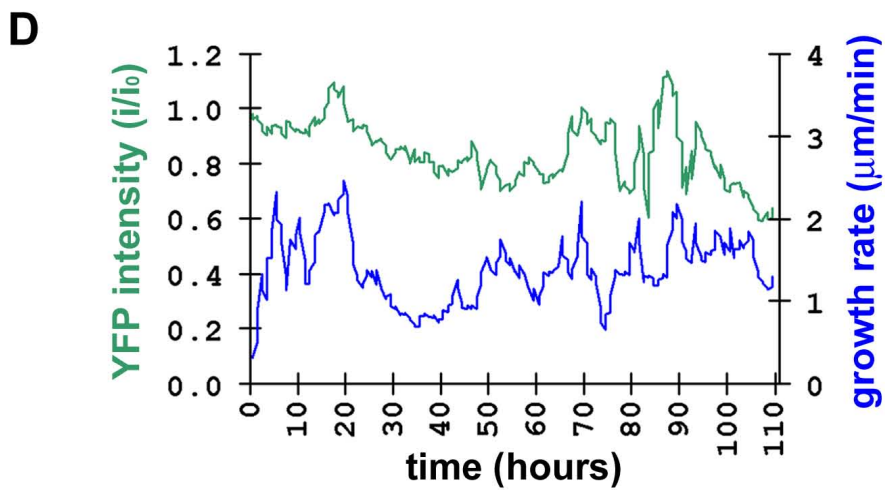

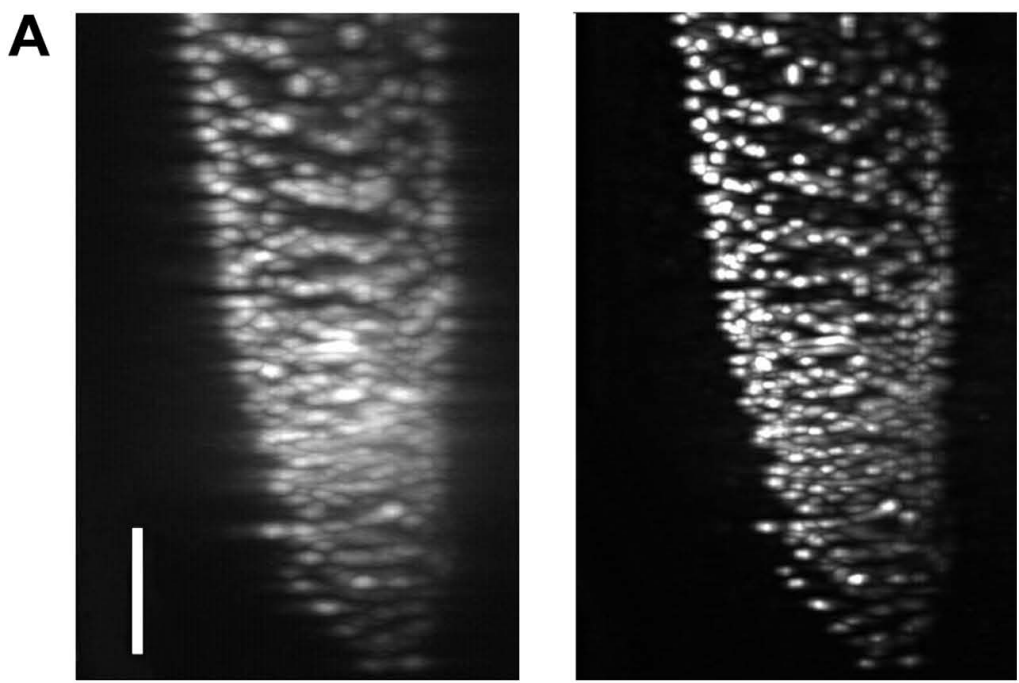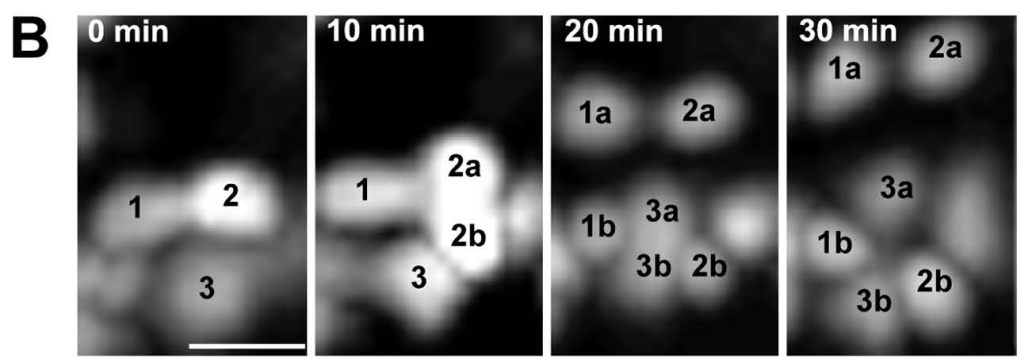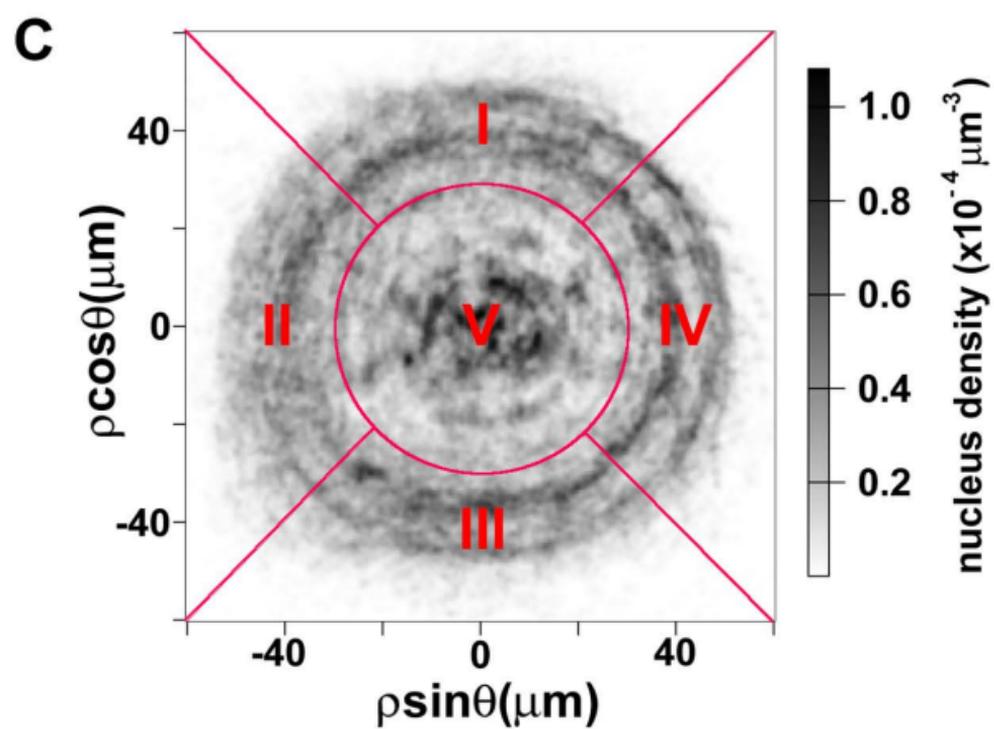

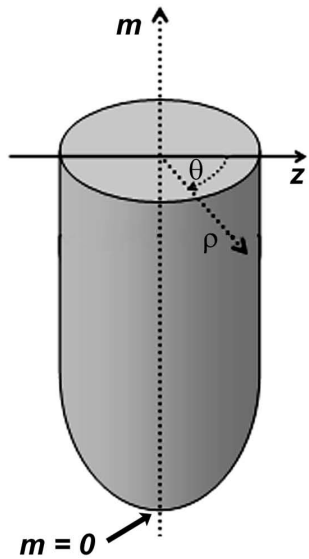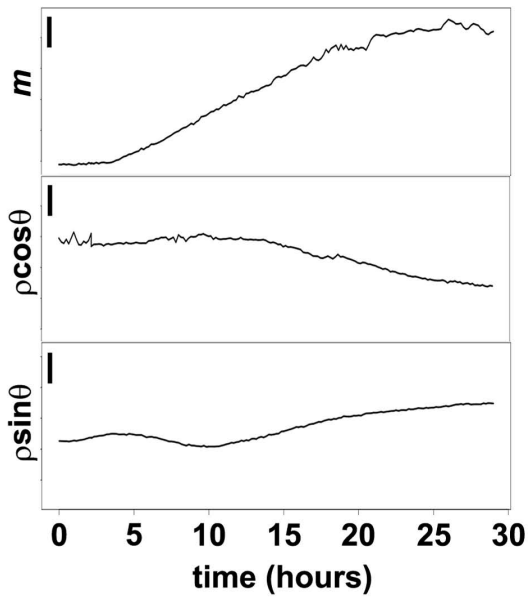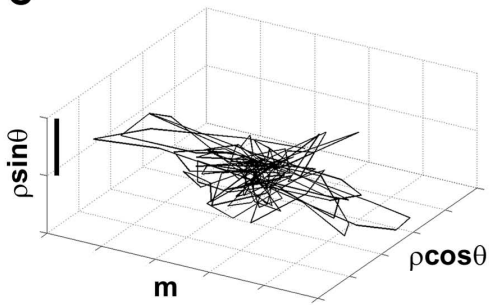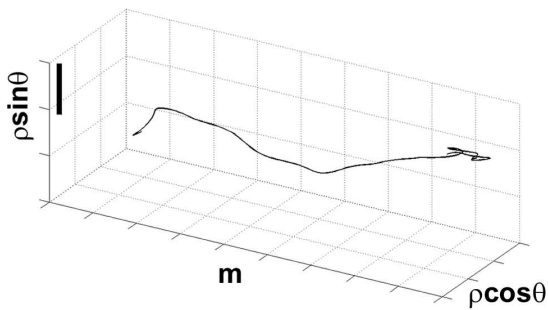

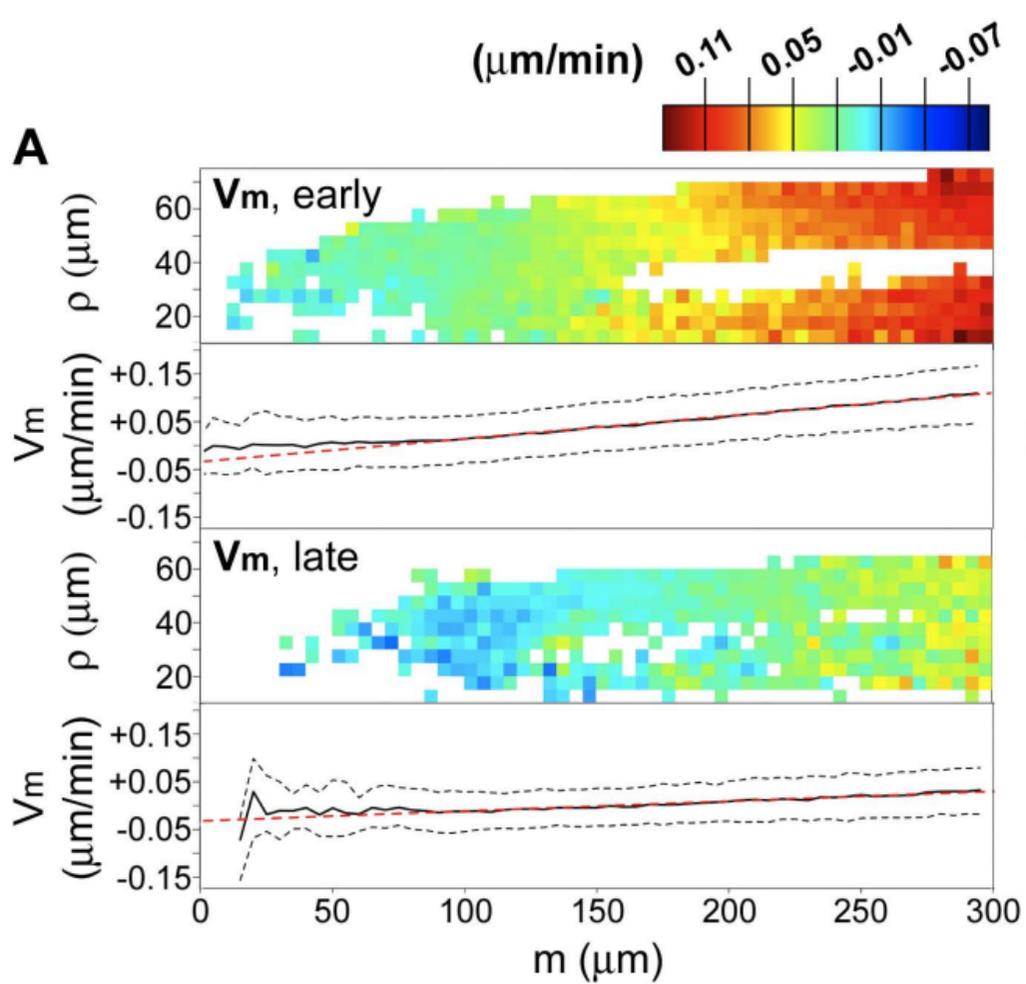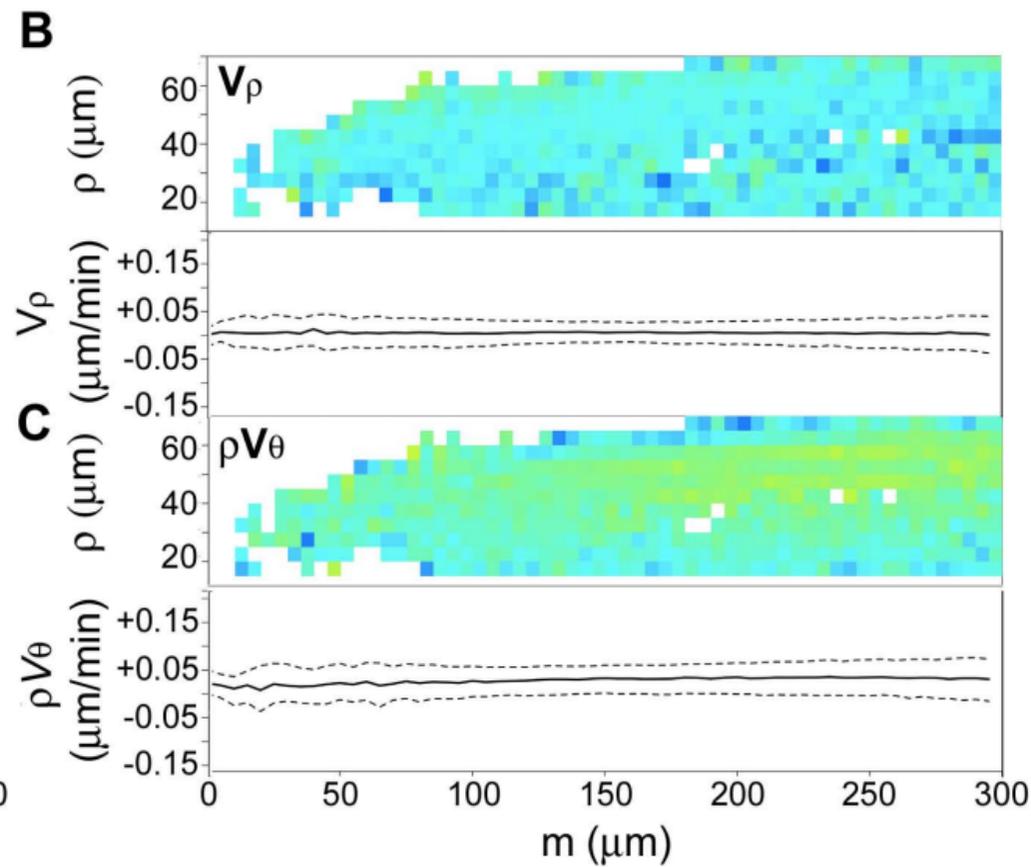

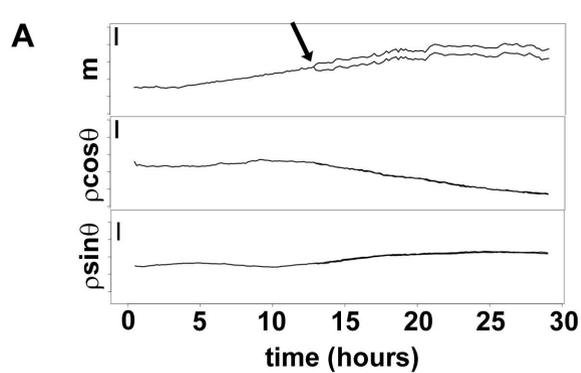
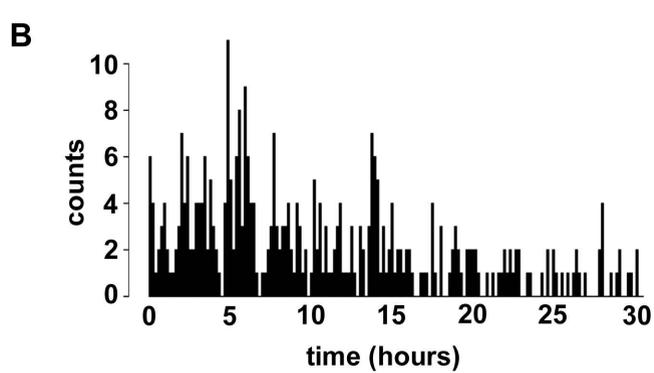
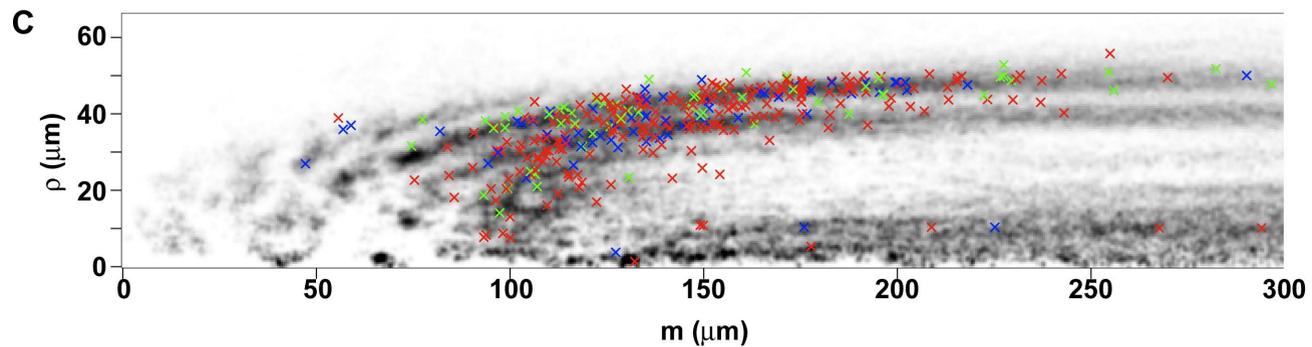
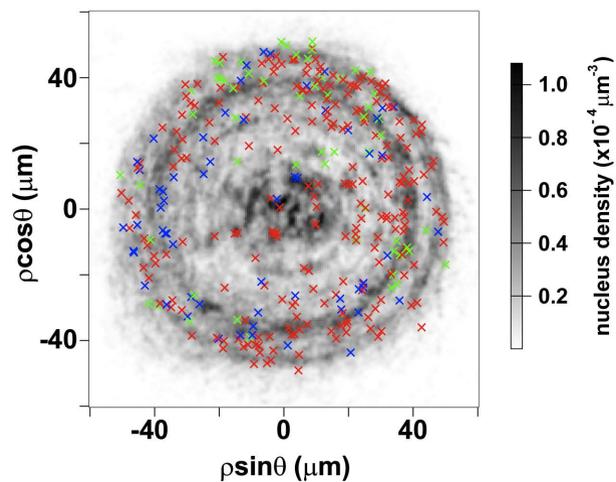
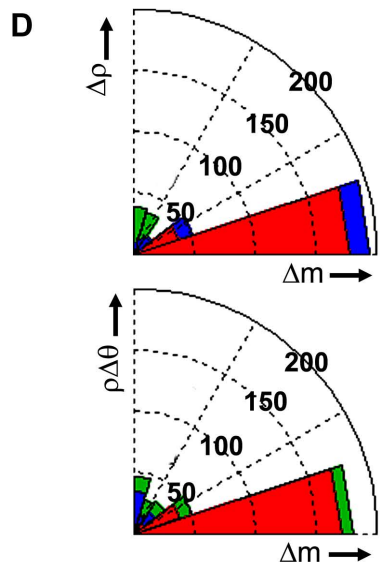

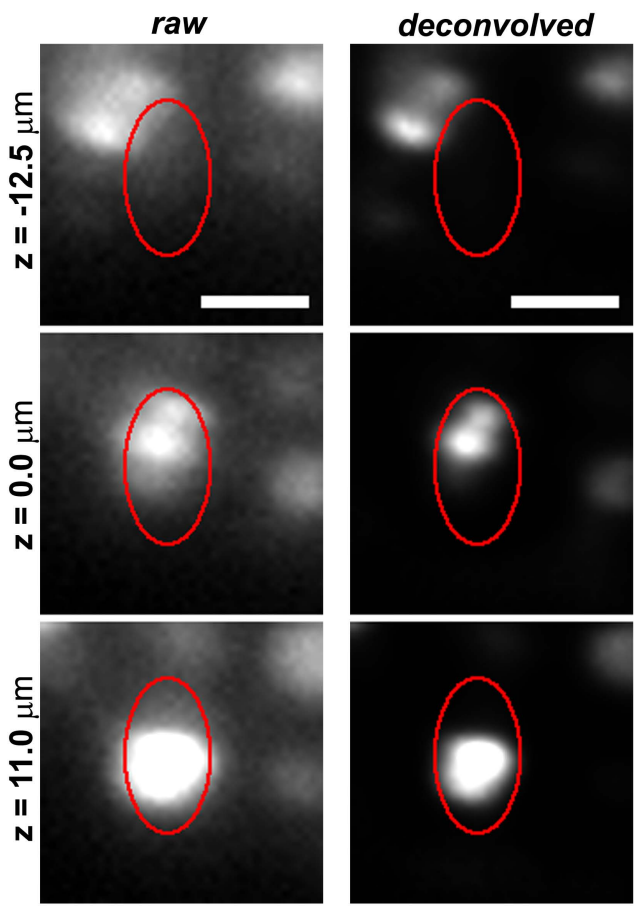

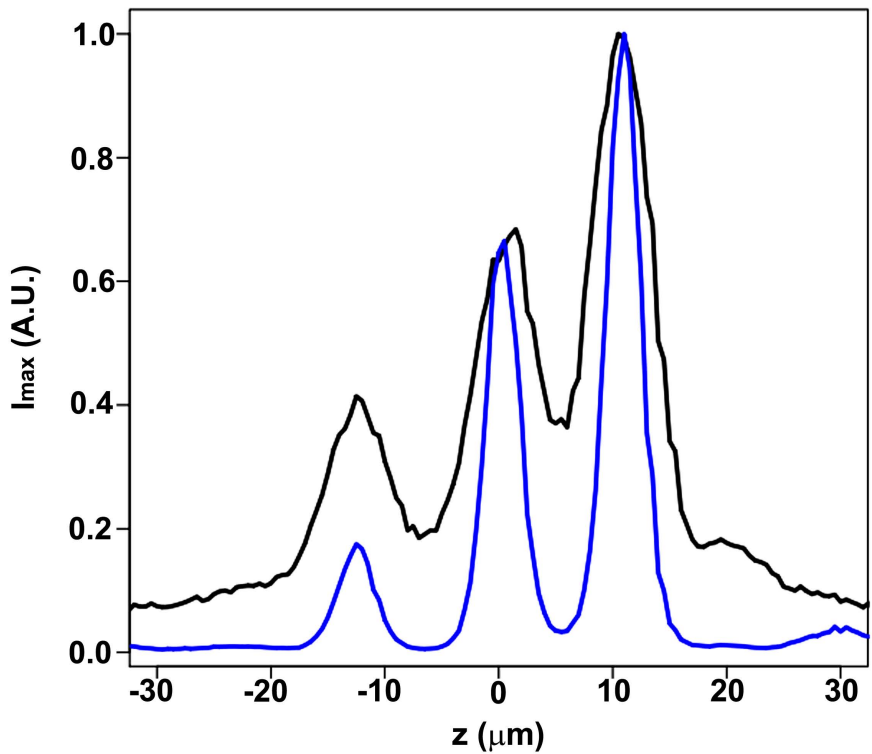

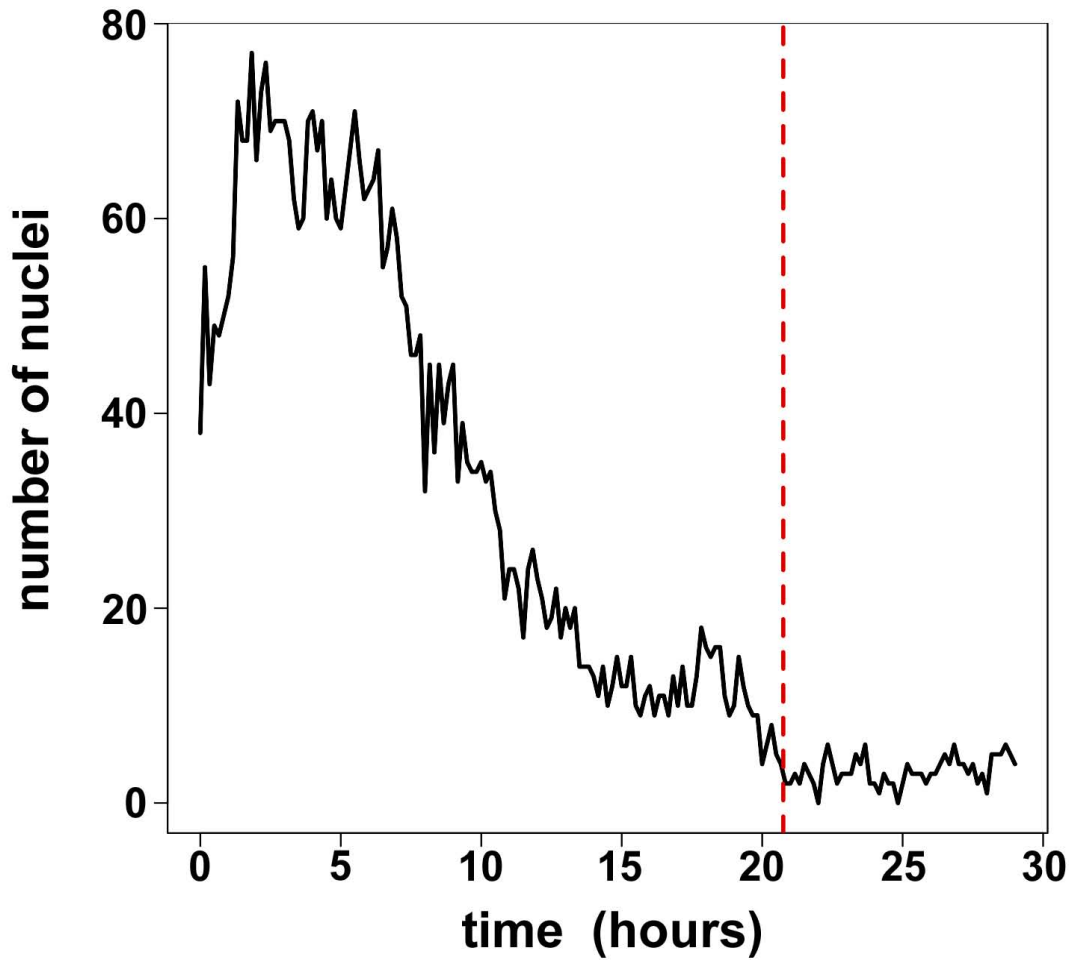

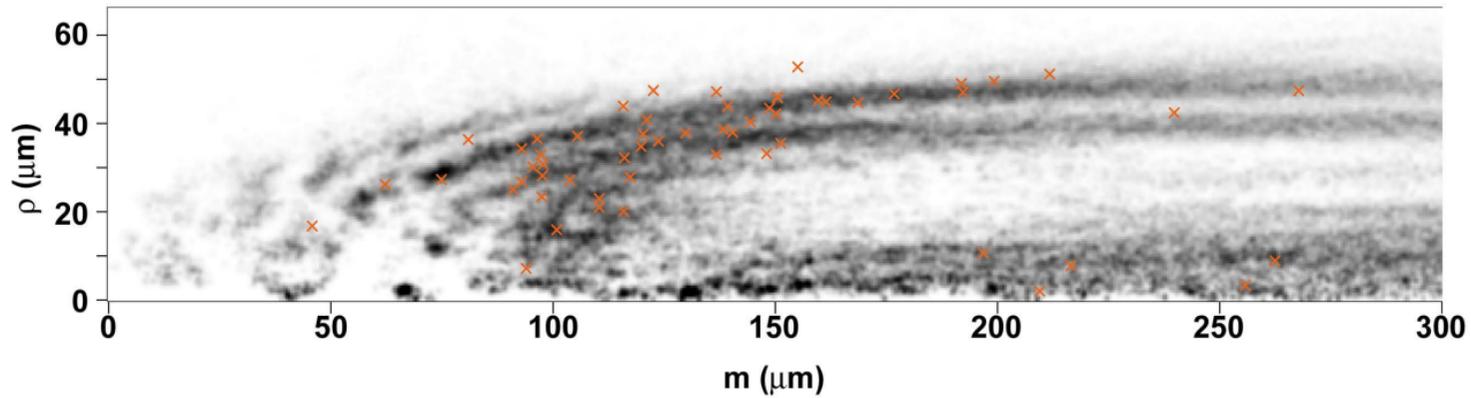

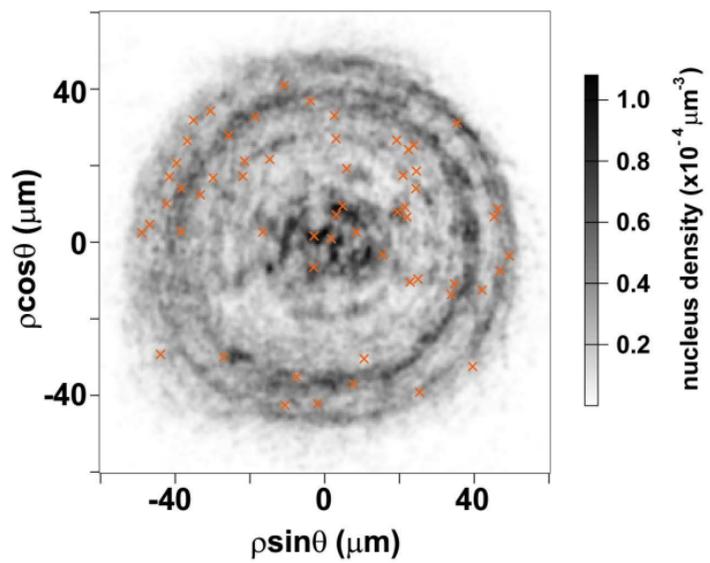

**A** 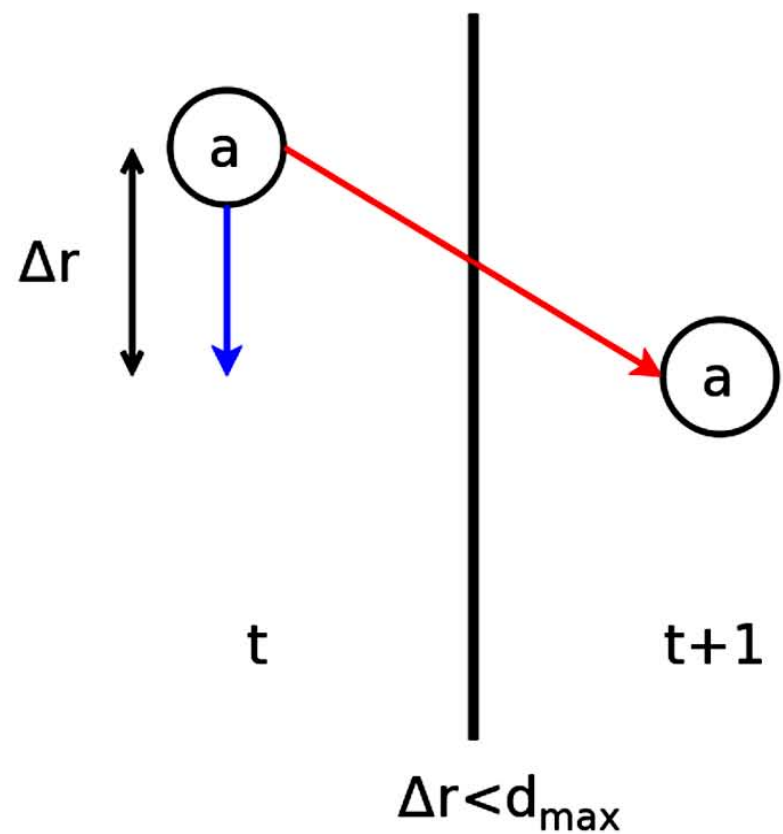

**B** 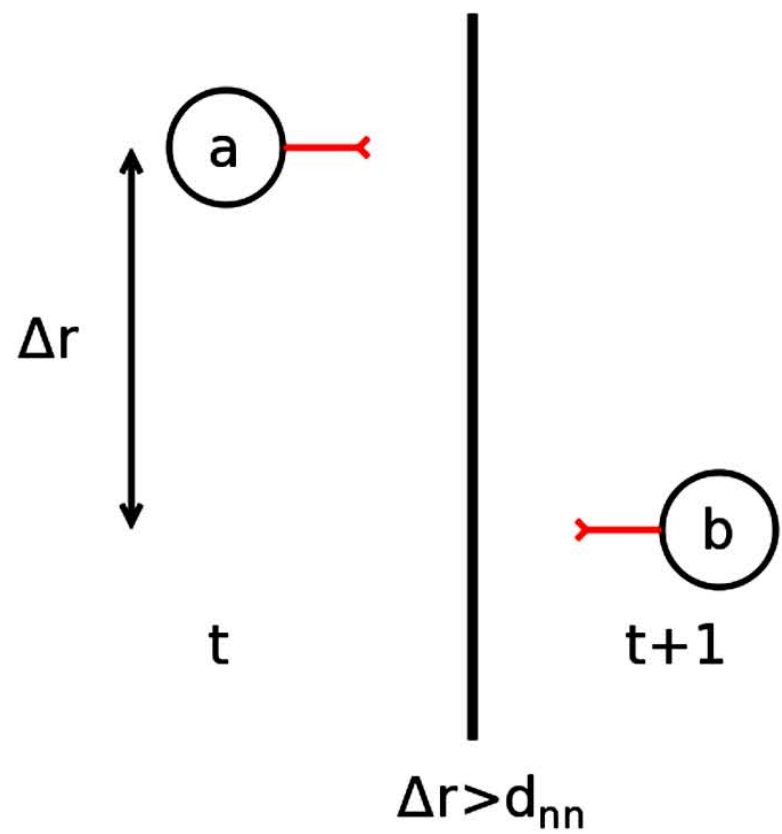

# Table S1

|     | *Segmentation* | | | *Tracking* | |
| --- | --- | --- | --- | --- | --- |
|     | N | FP (%) | FN (%) | N | Error range (%) |
| I   | 255 | 1.6 | 9.4 | 18 | 1.3 – 2.4 |
| II  | 238 | 3.8 | 9.2 | 16 | 3.9 – 5.6 |
| III | 220 | 11.4 | 5.9 | 15 | 2.3 – 3.8 |
| IV  | 239 | 1.7 | 5.0 | 19 | 1.4 – 2.4 |
| V   | 351 | 15.1 | 5.4 | 32 | 2.8 – 4.6 |
| **TOT** | **1303** | **7.3** | **6.9** | **100** | **2.4 – 3.9** |